\documentclass[%
 reprint,
 amsmath,amssymb,
 aps,
]{revtex4-2}

\usepackage{graphicx}
\usepackage{dcolumn}
\usepackage{bm}
\usepackage{mathtools}
\usepackage{caption}
\usepackage{subfig}
\usepackage{xcolor}
\usepackage[normalem]{ulem}
\makeatletter
\def\@makecaption#1#2{%
  \par\smallskip
  \begingroup
    \leftskip=0pt
    \rightskip=0pt
    \parindent=0pt
    \noindent #1.\ #2\par
  \endgroup
}
\makeatother

\usepackage{hyperref}

\def\>{\rangle}\def\<{\langle}
\def\togli#1{}

\begin{document}

\preprint{APS/123-QED}

\title{First-Click Time Measurements} 

\author{Mafalda Pinto Couto$^{1,3}$, Lorenzo Maccone$^2$, Lorenzo Catani$^3$ and Simone Roncallo$^2$}

 \affiliation{
1.~Universidade do Minho, Braga, Portugal}

\affiliation{2.~Dip.~Fisica A.Volta and INFN Sez. Pavia, University of Pavia, Via Bassi 6, I-27100 Pavia, Italy
}
 \affiliation{
 3.~INL - International Iberian Nanotechnology Laboratory, Avenida Mestre José Veiga s/n, 4715-330 Braga, Portugal
}

\date{\today}

\begin{abstract}
There are two distinct perspectives on the quantum time-of-arrival: one can ask for the probability that a particle is found at the detector at a given time, regardless of whether it was previously detected, or for the probability that the particle is detected there for the first time. In this work, we analyze the latter by constructing the time-of-arrival distribution conditioned on the particle not having been detected at earlier times --- the \textit{first-click distribution}. We work within the Page and Wootters formalism, where time is treated as a quantum observable, and introduce a memory mechanism that records the outcomes of successive detection attempts separated by the detector's finite time resolution. We apply this framework to a single Gaussian wave packet and to a superposition of two overlapping wave packets. We find that conditioning on non-detection redistributes probability toward earlier arrival times, producing narrower and sharper distributions compared with the standard unconditioned case. This effect persists in the presence of quantum interference, though coarser time resolutions broaden the distribution and shift it toward later times.

\end{abstract}

\maketitle

In textbook quantum mechanics, time is treated as an external classical parameter rather than as an observable like position or momentum. This asymmetry has motivated several attempts to formulate time within quantum theory itself~\cite{Stueckelberg, ZEH2009, Rovelli1996, PhysRevD.43.442, PageandWootters, 4Canadian, Cotler_2016, Yakir, sels2015thermodynamicstime, PhysRev.109.571, Feyman}. 
Among these, the Page and Wootters 
formalism \cite{Wootters, Page, PageandWootters} introduces time as a quantum observable associated with a clock system. This clock is defined as a quantum system distinct and independent from the system under study. In this framework, the combined state of system and clock is stationary with respect to any external time,  and temporal evolution emerges only through the correlations (entanglement) between system and clock. This approach also provides a general prescription for quantum measurements of time \cite{MacconeQTM}, enabling, in particular, the computation of time-of-arrival (TOA) distributions for particles at a given detector. While this framework provides a natural TOA distribution, it does not isolate the first detection event, which is often the experimentally relevant quantity in repeated monitoring settings. Different approaches to the TOA problem have been proposed. Within textbook quantum mechanics, these include time-of-arrival operators and their quantization \cite{Grot_1996,Delgado_1997,Aharonov21,Galapon_2004,Halliwell_2015,Galapon_2018}, detector-based approaches \cite{Echanobe_2008,PhysRevA.64.012501,PhysRevA.102.053705,Anastopoulos_2006,Halliwell_2009,Anastopoulos_2012}, and more general formulations of arrival-time distributions \cite{jurić2022arrivaltimegeneraltheory,Jurman_2021,Vona_2013,dhar2015quantumtimearrivaldistribution}. Other proposals arise from extensions or alternative formulations of quantum theory \cite{MacconeQTM,MacconeQT,kazemi2025arrivaltimeclassical,simoneroncallo,Brunetti2010}, as well as from Bohmian mechanics \cite{Das2019,PhysRevA.58.840,GaugeQuantumTime}. A thorough analysis of several of these approaches is presented in \cite{Roncallo_2023}.

In this paper, we compute the probability that a particle is detected at a given time, conditioned on it not having been detected at earlier times: the \textit{first-click distribution}. To achieve this, we introduce a memory system that records the outcomes -- detection and non-detection -- of successive detection attempts performed by the detector. These attempts occur at discrete time intervals determined by the finite time resolution of the detector. This construction allows us to define a global state that encodes the full detection history, from which the first-click probability distribution can be derived.  
We analyze the first-click distribution numerically for some representative initial states: a single Gaussian wave packet and a superposition of two overlapping Gaussian wave packets. 
We perform a comparison between these results and those obtained from the framework of \cite{MacconeQTM}, which in this paper will be referred to as the ``memoryless framework''. 
We show that conditioning on no earlier detections, or \textit{clicks}, modifies the TOA distribution relative to the memoryless case, and how this modification depends on the detector time resolution. We find that conditioning redistributes probability toward earlier arrival times, yielding narrower distributions whose shape depends on the detector time resolution. Each null detection constitutes a measurement that updates the particle's state, an effect entirely absent in the memoryless framework.

The paper is structured as follows. In Section \ref{sec:level1} we review the Page and Wootters formalism \cite{Wootters, Page, PageandWootters} and its extension to the TOA problem \cite{MacconeQT, MacconeQTM}. In Section \ref{sec:level2}, we develop a framework to compute the first-click distribution using a memory mechanism. In Section \ref{sec:level3} we present numerical results for two representative cases.  
In Section \ref{sec:level4}, we discuss the implications of our results and outline possible directions for future research. 

\section{\label{sec:level1} Background}
In this section we review the Page and Wootters formalism and how it describes time measurements.\par

\subsection{Page and Wootters formalism}
In this formalism, the temporal evolution of a physical system $S$ is described through its correlations with an ancillary system $T$, which acts as a clock with Hilbert space $\mathcal{H}_T$. For example, a good clock is a 1d particle whose position $|t\rangle$ tells the time and whose Hamiltonian $H_T$ is the particle's momentum. The system under consideration is described on a Hilbert space $\mathcal{H}_S$ and evolves under a Hamiltonian $H_S$.  \par

The total closed system is described on the composite Hilbert space $\mathcal{H}\coloneqq \mathcal{H}_T\otimes\mathcal{H}_S$. Its global state, denoted by $|\Psi\rangle$, is taken to be an eigenstate of the total Hamiltonian $\hat{H}=\hat{H}_T\otimes\mathbb{I}_S+\mathbb{I}_T\otimes\hat{H}_S$, where the clock and system are assumed not to interact. The global state is stationary with respect to external time, but clock and system are entangled so that effective temporal evolution of the system $S$ emerges from its correlations with the clock. 
Indeed, the conditional state of the system $S$ at clock reading $t$ is obtained by projecting the global state $|\Psi\rangle$ onto the clock state $|t\rangle_T$. Explicitly, $|\psi (t) \rangle _S = \hspace{0.1cm}_T \langle t | \Psi \rangle$, where $|t\rangle_T$ denotes a "position" eigenstate of the clock time operator $\hat{T}$. \par 

In the time basis, the global state can therefore be expanded as
\begin{equation} \label{eq:funcintegral}
    |\Psi \rangle = \int dt |t\rangle_T \otimes | \psi (t) \rangle_S.
\end{equation}

\noindent Since this expression involves an integral over an unbounded time domain, it is not normalizable in the usual sense. A standard regularization is to restrict the integral to a finite interval, writing  
\begin{equation}
    |\Psi\rangle = \frac{1}{\sqrt{T}} \int_{-\frac{T}{2}}^{\frac{T}{2}} dt |t \rangle_T |\psi (t) \rangle_S,
\end{equation}

\noindent and then considering the limit in which $T$ is much larger than all other relevant timescales in the problem. \par

By choosing the clock Hamiltonian equal to its momentum, it is canonically conjugate to the time operator $\hat T$, so that it acts as $-i\hbar \partial_t$ in the time representation. One then recovers the Schr\"odinger equation for the conditional state of the system, and hence the usual quantum dynamics.

\subsection{Quantum Measurements}

Measurement processes within the Page and Wootters formalism can be described \cite{MacconeQT} with the von Neumann formulation  of measurement \cite{vonNeumann+2018}. Namely one introduces an ancillary memory system $M$ that is dynamically coupled (instantaneously, ideally) to the system to be measured and stores the measurement outcome. So now the system $S$  comprises both the measured system $Q$ and the memory $M$ 
\begin{align}
	\hat{H}_S (t) = \hat{H}_Q(t) + \delta (t-t_1) \hat{h}_{QM} ,
\end{align}
\noindent where $\hat{h}_{QM}$ represents the interaction Hamiltonian between $Q$ and $M$ at the instant $t_1$. Here, the time $t$ refers to the clock observable.

Prior to the measurement, $M$ is in a ready state $|0\rangle_M$, and upon detection, it transitions to a state $|a\rangle_M$ that encodes the measurement outcome $a$. This process is modeled as an ideally instantaneous von Neumann-type interaction at time $t_1$, resulting in the transformation
\begin{equation} \label{eq:kraus}
    |\psi (t_1) \rangle_Q \otimes |0\rangle_M \rightarrow \sum_a \hat{K}_a |\psi (t_1) \rangle_Q \otimes |a\rangle_M,
\end{equation}

\noindent where $\{ \hat{K}_a \}$ are Kraus operators associated with the measurement outcomes. E.g.~for ideal projective measurements, $K_a=|a\>\<a|$, a projector on the observable eigenstate $|a\>$, so that we obtain the usual post-measurement state $\sum_a \psi_a|a\rangle_Q \otimes |a\rangle_M$, with $\psi_a=\<a|\psi\>$.\par 
Accordingly, for a single measurement performed at time $t_1$, the global state becomes
\begin{align} \label{eq:globalstateone}
    |\Psi \rangle = & \int_{-\infty}^{t_1} dt |t\rangle_T \otimes |\psi (t) \rangle_Q \otimes |0\rangle_M \nonumber \\
    + & \int_{t_1}^\infty dt |t\rangle_T \otimes \sum_a \hat{U}_Q(t-t_1) \hat{K}_a |\psi (t_1) \rangle_Q \otimes |a\rangle_M,    
\end{align}
\noindent where $\hat{U}_Q(t-t_1)$ denotes the unitary operator governing the free evolution of $Q$ after the measurement. \par

This formulation extends naturally to sequences of measurements performed at distinct times.

\subsection{Time Of Arrival Problem}

In \cite{MacconeQTM}, the Page and Wootters formalism is applied to the time-of-arrival  problem for a one-dimensional particle and a detector associated with a spatial region $D$. \par

To obtain the arrival-time distribution, one introduces a TOA projection-valued measure (PVM) defined by 
\begin{equation} \label{eq:POVM}
    \forall t: \Pi_t \equiv |t \rangle\langle t| \otimes P_d ,
\end{equation}

\noindent where $P_d = \int _D dx |x \rangle \langle x|$ projects onto the detector's spatial region $D$, and its completion $ \mathbb{I}-\int dt \Pi _t$. The operators $\{\Pi_t\}$ correspond to detection events at time $t$.

With this definition, the joint probability that the particle is found in $D$ and the clock records a time $t$ is  
\begin{equation} \label{eq:firstjointprob}
    p(t,x \in D) = Tr \left[ |\Psi\rangle \langle\Psi | \Pi_t \right] = \frac{1}{T} \int _{x\in D} dx \left\lvert \psi (x|t) \right\lvert ^2,
\end{equation}

\noindent where $\psi(x|t)=\langle x|\psi(t)\rangle$ is the system wavefunction conditioned on the clock reading $t$. \par 
Applying Bayes' theorem, the time-of-arrival distribution conditioned on detection in $D$ is
\begin{equation} \label{eq:timedistribution}
    p(t|x \in D) =  \frac{p(t,x \in D)}{p(x \in D)}  = \frac{ \int _{x\in D} dx \left\lvert \psi (x|t) \right\lvert ^2}{\int_T dt \int _{x\in D} dx \left\lvert \psi (x|t) \right\lvert ^2}.
\end{equation}

\par

\noindent This construction yields the standard memoryless distribution. It will serve as the reference distribution against which the distribution introduced in the next section is compared.

\section{\label{sec:level2} First Click }

The TOA distribution obtained in Eq.~(\ref{eq:timedistribution}) allows for the possibility of multiple detections of the same particle, referred to as click time measurements, or simply \textit{clicks}. In many experimental situations, however, the relevant quantity is the probability that the detector clicks for the first time at a given instant.   \par

To describe this situation, we must monitor the particle's arrival throughout the entire duration of the experiment. Since detection events are identified by the detector, our goal is to store the outcomes of all detection attempts in a memory mechanism. Ideally, this monitoring would be continuous. In practice, however, any detector has a finite time resolution, so successive detection attempts are separated by a nonzero time interval $\delta t$. This discreteness must therefore be incorporated into the formalism. \par 
By introducing a memory system that records the outcome of each measurement attempt, we can compute the first-click probability distribution. \par

\subsection{Experimental Setup}

The experiment is conducted over a total time $T$, starting at $-\frac{T}{2}$ and ending at $\frac{T}{2}$. 
The detector's time resolution is $\delta t = \frac{T}{n}$, where $n$ represents the total number of times the detector can, in principle, probe the particle during the experiment. The corresponding times are
\begin{align}
    \bigg\{-\frac{T}{2}, \ -\frac{T}{2} + \delta t, \ -\frac{T}{2} + 2 \delta t, \ ..., \ \frac{T}{2} - \delta t  \bigg\} .
\end{align}
 At each detection attempt time, the system is probed and the arrival information is recorded. Even when the detector does not click, it still provides information about the system, specifically that the particle lies outside the detector's spatial region. This non-click introduces non-negligible effects on the system's evolution: the particle undergoes an instantaneous transformation characterized by the appropriate Kraus operators. \par

The two primary subsystems are the clock, denoted by $T$, and the system, denoted by $S$. The system $S$ consists of the particle system, $Q$, and a memory system, $M$, which stores the detection attempts outcomes in $n$ distinct memory states. More precisely, we associate one memory subsystem $M_i$ with each detection attempt time $t_i$, for $i=0,\dots,n-1$. The detector occupies a spatial region $D$ of length $\delta L$. \par 

The free evolution of the particle over a time interval $\Delta t$ is described by the unitary operator $\hat{U}_Q(\Delta t)$
\begin{equation}
    \hat{U}_Q(\Delta t) = e^{-i\Delta t \hat{H}_Q /\hbar},
\end{equation}
\noindent where $\hat{H}_Q$ is the Hamiltonian of $Q$. \par

At the $i^{th}$ detection attempt, for $0\leq i \leq n-1$, the clock reads $t_i$, where $t_0 = -\frac{T}{2}$. The corresponding memory state is initially in the state $|0\rangle_{M_i}$. If the detector clicks at time $t_i$, the memory state is updated to $|1\rangle_{M_i}$, otherwise it remains in the state $|0\rangle_{M_i}$. \par \bigskip

Since our objective is not to determine the value of a specific observable but rather to establish whether a click occurs, the Kraus operators are taken to be the projectors onto the click and non-click eigenspaces determined by the detector's spatial domain
\begin{align} \label{eq:KrausOperator1}
    \hat{K}_1 & =\int_D dx \ |x \rangle_Q\langle x| ,\\
    \hat{K}_0 & = \int_{x \notin D} dx\ {|x\rangle_Q\langle x|  } . \label{eq:KrausOperator0}
\end{align} 
\noindent Here, $\hat{K}_1$ corresponds to detection within $D$, while $\hat{K}_0$ corresponds to non-detection.

\subsection{Global State Derivation}

We now construct the global state of the clock-system-memory composite, encoding the full detection history of the experiment. \par 

At each detection attempt time $t_i$, with $i \in \{0, \dots, n-1\}$, the probing of the particle and the storage of the corresponding outcome are represented by the evolution
\begin{align}
    \hat{K}_1 \otimes |1\rangle_{M_i}\<0| + \hat{K}_0 \otimes |0\rangle_{M_i}\<0|,
\end{align}
which simultaneously updates the state of the particle and records the outcome in the $i^{\text{th}}$ memory. \par

After the $i^{\text{th}}$ detection attempt, the particle evolves freely from $t_i$ to a generic time $t \in [t_i,t_{i+1}]$, yielding the factor $\hat{U}_Q(t - t_i)$. All previous detection attempt outcomes and their effects on the state are encoded by the sequence of alternating unitary evolution and measurement operations. This results in a time-ordered product of the form
\begin{align}
    \prod_{j=1}^i \hat{U}_Q(\delta t)\big( \hat{K}_1 \otimes |1\rangle_{M_{i-j}}\<0| + \hat{K}_0 \otimes |0\rangle_{M_{i-j}}\<0| \big),
\end{align} 

\noindent where between consecutive detection attempts the particle evolves over a fixed interval $\delta t = \frac{T}{n}$. The last detection attempt occurs at $t_{n-1}$ and is followed by free evolution up to $t_n = \frac{T}{2}$, the final instant of the experiment. \par

We now combine these ingredients into a single expression for the global state,
\begin{widetext}
\begin{align} \label{eq:globalstatefinal}
    |\Psi \rangle & = \frac{1}{\sqrt{T}} \Bigg[ \sum_{i=0}^{n-1} \int_{t_i}^{t_{i+1}} dt \ |t\rangle _T\otimes \:\hat{U}_Q\left( t - t_i \right) \bigg( \hat{K}_1 \otimes |1\rangle_{M_i}\<0| + \hat{K}_0 \otimes |0\rangle_{M_i}\<0| \bigg) \nonumber   \\ 
    &\times  \prod_{j=1}^i \hat{U}_Q( \delta t ) \bigg( \hat{K}_1 \otimes |1\rangle_{M_{i-j}}\<0| + \hat{K}_0 \otimes |0\rangle_{M_{i-j}}\<0| \bigg)  \Bigg]
     | \psi ( t_0 ) \rangle _Q \bigotimes_{j=1}^{n}|0\rangle_{M_j} ,
\end{align}
\end{widetext}
where we coherently sum over all possible time intervals $[t_i,t_{i+1}]$, weighted by the clock states $|t\rangle_T$. Each term in the sum corresponds to the integral on the corresponding interval in which the clock is found. For a given $i$, the expression inside the integral has a clear structure: the product term encodes the full history of measurement attempts from $t_0$ up to $t_i$; the Kraus operators acting on $M_i$ represent the detection attempt performed at time $t_i$; the unitary $\hat{U}_Q(t - t_i)$ describes the free evolution from $t_i$ to the continuous time $t$; and the remaining memory subsystems $M_{j>i}$ stay in their initial state $|0\rangle$, since no measurement attempt has yet been performed on them.

This construction explicitly encodes the full sequence of detection attempt outcomes and their back-action on the system. At each measurement attempt, the state is updated via the appropriate Kraus operator, the outcome is stored in a memory qubit, and the particle subsequently evolves unitarily until the next measurement attempt.

\subsection{Time Of Arrival Distributions}
To obtain the probability distribution of Eq.~\ref{eq:timedistribution}, the global state was projected onto the appropriate subspaces of the composite Hilbert space, using the PVM introduced in Eq.~(\ref{eq:POVM}).\par
In the present case, however, we impose an additional constraint: we project the state onto the subspace corresponding to detection at time $t_f$ ( where $f$ stands for ``first'') together with no detection at any earlier time. This restriction is enforced by specifying the relevant memory states: 
the memory state at time $t_f$ must be $|1\rangle_{M_f}$, while those associated with earlier times must be $|0\rangle_{M_{i}}$, for $i < f$. No condition is imposed on the memory subsystems associated with times later than $t_f$. \par

The PVM describing the \textit{first-click} is thus defined by 
\begin{align} \label{eq:pvm}
    &\Pi_{t_f} =  
    |t_f \rangle_T\langle t_f| \   \bigotimes_{i=0}^{f-1} \bigg( |0\rangle _{M_i}\langle0| \bigg) \otimes |1\rangle_{M_f}\langle1| \otimes P_d, 
\end{align} 
with $f = {t_f}/{\delta t}$, and its completion $\mathbb{I}-\int dt \Pi_{t_f}$. This projective measure has outcome $t_f$ when the detector first clicks at time $t_f$.

Applying the Born rule to Eq.~(\ref{eq:globalstatefinal}) and Eq.~(\ref{eq:pvm}), we obtain the joint probability
\begin{align}\label{eq:jointdistr}
    p(t_f, x\in D, |0\rangle_{M_i} \forall i<f) = Tr[|\Psi\rangle\langle \Psi| \Pi_{t_f}] = \nonumber \\ 
    = \frac{1}{T} \int _D dx |_Q\langle x |\psi_c(t_f) \rangle_Q |^2 ,
\end{align}
where $|\psi_c(t_f)\rangle_Q$ is the particle state conditioned on the detector clicking at time $t_f$ and not clicking at earlier times, namely    
\begin{align} \label{eq:conditionedstate}
    & |\psi_c \left( t_f \right) \rangle_Q  = \hat{K}_1 \left( \prod_{j=1}^{f-1} \hat{U}_Q\left( \delta t \right) \ \hat{K}_0 \right) |\psi \left( t_0 \right) \rangle_Q .
\end{align} 

From the joint probability, we can now derive the first-click distribution via Bayes' rule
\begin{align} \label{eq:ftimedist}
     & p(t_f|x\in D, |0\rangle_{M_i} \forall \ i<f) = \nonumber \\ 
    &= \frac{\int _D dx |_Q\langle x |\psi_c(t_f)\rangle_Q|^2}{\sum_{i=0}^{n-1}\int_{-\frac{T}{2}+\frac{T}{n}i}^{-\frac{T}{2}+\frac{T}{n}(i+1)}dt\ \int _D dx |_Q\langle x |\psi_c(t_f)\rangle_Q|^2} .
\end{align}

Notably, in addition to the usual time integral, the denominator includes also a discrete sum over the allowed detection attempt times, reflecting the detector's finite temporal resolution. \par

As expected, the obtained first-click distribution closely mirrors the TOA distribution given in Eq.~(\ref{eq:timedistribution}): it retains the same basic structure, but incorporates the {\em crucial} additional requirement that no click occurs at any time earlier than $t_f$, the ``first-click'' time.

\section{Results \label{sec:level3}}

In this section, we numerically compare the predictions of the proposed framework, which will be referred to as the first-click framework, with those of the memoryless framework. 
In particular, we consider two representative cases: a single Gaussian wave packet and a superposition of two different-speed Gaussian wave packets overlapping at the detector. The corresponding TOA distributions are obtained from Eq.~(\ref{eq:timedistribution}) for the memoryless case and Eq.~(\ref{eq:ftimedist}) for the first-click case. \par

Throughout this section, we denote by $p_{ML}$ the probability distribution in the memoryless framework and by $p_{FC}$ the probability distribution in the first-click framework.   
In both cases, we do not introduce any specific model for the particle-detector interaction. This isolates the effect of the conditioning itself, allowing for a direct comparison between the two approaches.\par \bigskip

Consider a Gaussian wave packet $|\psi\>=\int dx\:\psi(x,{t})|x\>$ with $|x\>$ the position eigenstate and with initial position $x_0$, momentum $p_0$ and standard deviation $\sigma_0$
\begin{align}
    \psi (x,0) = \frac{1}{(\pi \sigma_0^2 )^{1/4}} e^{\frac{-(x-x_0)^2}{2 \sigma_0^2}+ip_0(x-x_0)/\hbar}.
\end{align}
Its time evolution is governed by the free Hamiltonian $\hat{H}=\hat{p}^2/2m$, where $m$ is the particle mass. The wave packet evolves according to 
\begin{align} \label{Eq:WaveEvolve}
    \psi (x,t) = \frac{e^{-\frac{(x-x_0-p_0t/m)^2}{2\sigma_0^2 (1+i\hbar t/m \sigma_0^2)}} e^{\frac{ip_0}{\hbar}\left(x-x_0-\frac{p_0 t}{2m} \right)}}{\sqrt{\sqrt{\pi} \left( \sigma_0 + \frac{i \hbar t}{m \sigma _0}\right) }} .
\end{align}
In all simulations, the particle is initially prepared in the ground state of a harmonic trap with frequency $\omega$. We use units of time, length and energy, respectively given by $t_0 = 1 /\omega$, $l_0 = \sqrt{\hbar/m\omega}$ and $E_0 = \hbar \omega$. In what follows, we set $\hbar = m = 1$. The initial standard deviation of the packet is fixed to $\sigma _0 = 1 l_0$. \par \bigskip

To efficiently evaluate the numerical time evolution required to compute $p_{FC}$, we employ a fast Fourier transform (FFT) algorithm, exploiting the fact that the particle's evolution can be written as a convolution with the free-space propagator \cite{Shankar}, defined as
\begin{align} \label{Eq:EvolveU}
    \hat{U}(t) = \int _{-\infty}^\infty  dp\:|p\rangle\langle p| e^{-ip^2 t / 2m\hbar}.
\end{align}
\noindent This procedure significantly reduces the large computational cost of simulating Eq.~(\ref{eq:ftimedist}). Using the convolution theorem, we write 
\begin{align}
    \{u*v\}(x) = \mathcal{F}^{-1}\{ U\cdot V \} = \mathcal{F}^{-1}\{\mathcal{F}\{u\}(f) \cdot \mathcal{F}\{v\}(f)\},
\end{align}
where $u(x)$ and $v(x)$ are two functions with Fourier transforms $U$ and $V$, respectively. Here $\mathcal{F}$ denotes the Fourier transform and $\mathcal{F}^{-1}$ its inverse. This allows the required convolutions to be evaluated through simple multiplications in Fourier space, substantially reducing the computational cost. \par

Care is required when implementing FFT methods, since the choice of the spatial domain strongly affects the resulting distributions. In particular, if the spatial and momentum grids are not chosen large enough to contain the relevant support of the wave packet, numerical artifacts may appear. 
In addition, FFT algorithms assume periodic boundary conditions, which can lead to artificial wrap-around effects: the tail of the wave packet can re-enter the computational domain and be incorrectly interpreted as a new arrival. We mitigate this issue by introducing zero-padding at the boundaries, ensuring that the wavefunction propagates without spurious periodic contributions. \par

\subsection{Gaussian wave packet}

We first compare the two frameworks using a single Gaussian wave packet propagating from left to right. Its evolution is described by Eq.~(\ref{Eq:WaveEvolve}) in the memoryless framework and by Eq.~(\ref{Eq:EvolveU}) in the first-click framework. \par 
The detector occupies the spatial region $D = [a,b]$, where $a$ and $b$ denote the beginning and the end of the detector. Within the memoryless framework, we consider two detector configurations. The first is a finite-size detector, covering the full region $D$, such that in Eq.~\ref{eq:POVM} the projector is given by $P_d = \int_D dx |x\rangle\langle x|$. The second is a point-like detector, modeled as an idealized detector with negligible spatial extent and placed at the left boundary of $D$, corresponding to $P_d = |a\rangle\langle a|$. The latter configuration allows for a more direct comparison between the two frameworks. The first-click framework is analyzed assuming an ideal detector with the finest available time resolution, $\delta t = t_0$. \par \bigskip 

\begin{figure}[t!]
    
    \includegraphics[width=1\linewidth]{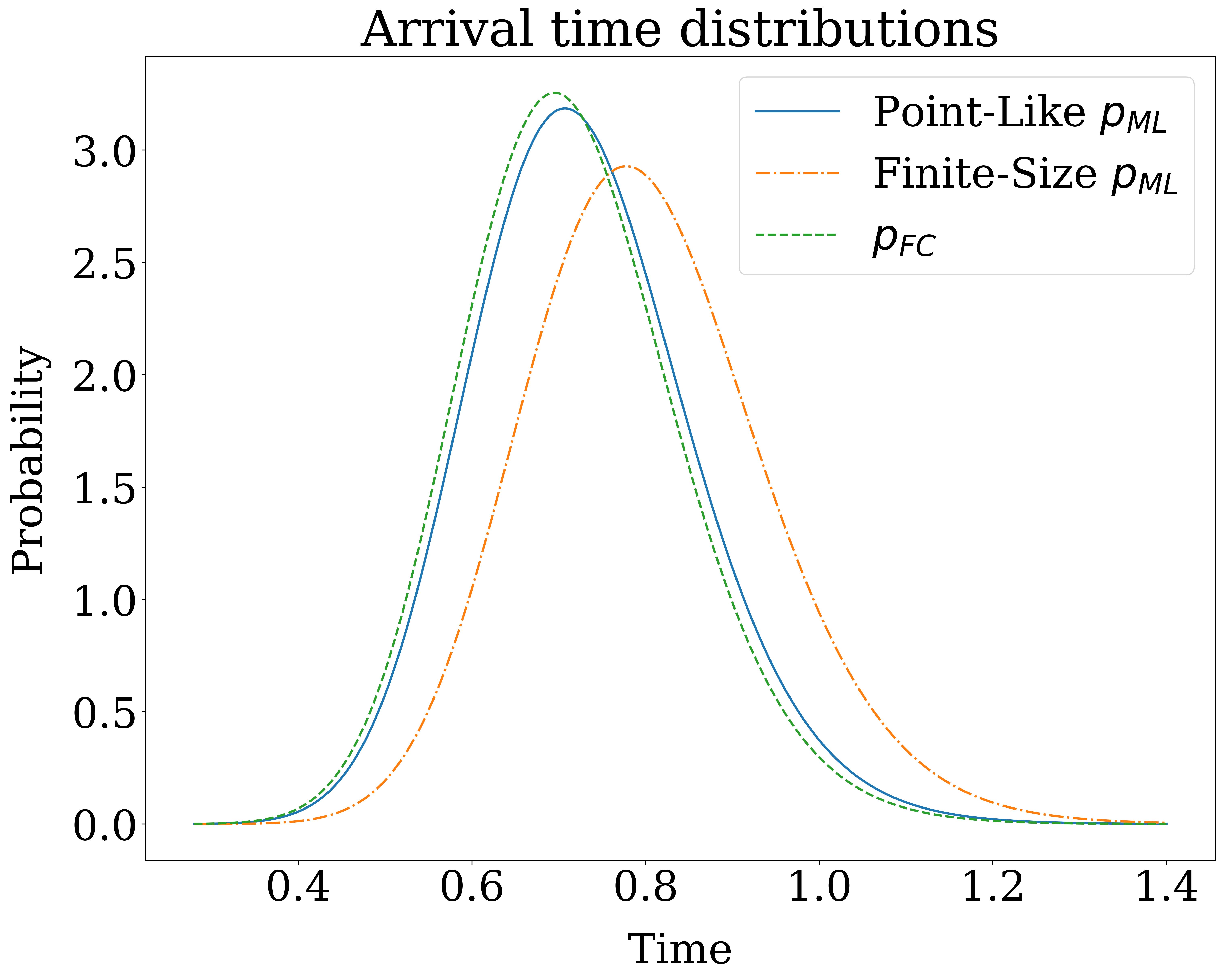}
    \caption{Time-of-arrival distributions for a single right-moving Gaussian wave packet propagating toward the detector. The packet parameters are $x_0 = 5l_0$ and $p_0 = 7\hbar/l_0$, corresponding to its initial average position and momentum, respectively. The detector’s time resolution is $\delta t = t_0$ and it extends between $a=10 l_0$ and $b=11 l_0$. The \textit{finite-size $p_{ML}$} is broader and exhibits a lower peak amplitude, consistent with its extended spatial range. The \textit{$p_{FC}$} distribution is narrower and shifted toward earlier times due to the conditioning on non-detection.} 
    \label{image:GaussianWavePacket}
\end{figure}

As shown in Fig.~\ref{image:GaussianWavePacket}, the finite-size $p_{ML}$ exhibits a broader and lower-amplitude distribution in comparison to the point-like $p_{ML}$, reflecting the extended spatial range of the detector. The temporal shift between the two $p_{ML}$ distributions arises from the different detector positions. In particular, this offset would vanish if the point-like detector were placed at the center of $D$, namely at position $x = (a+b)/2$. \par 

By contrast, $p_{FC}$ is shifted toward earlier times relative to both $p_{ML}$ distributions, while also being narrower and exhibiting a higher peak amplitude. This difference cannot be accounted for by a simple rescaling or normalization, but instead reflects a genuine reshaping of the distribution. \par 

This behavior originates from the conditioning imposed during the evolution. At each time step, the no-click requirement progressively suppresses the leading edge of the wave packet by removing amplitude from the detector region. The remaining state therefore becomes increasingly biased toward earlier arrival times, resulting in a redistribution of the probability weight and a sharpening of the distribution. {This interpretation becomes clear when the point-like detector is placed at the left boundary of the finite-size detector and the difference is still evident. Without this reference configuration, it would be difficult to disentangle the effects due to conditioning from those arising purely from the detector’s spatial position, and hence to see that the first-click framework cannot be identified with a point-like detection model.}

\begin{figure}[t!] 
    \includegraphics[width=1\linewidth]{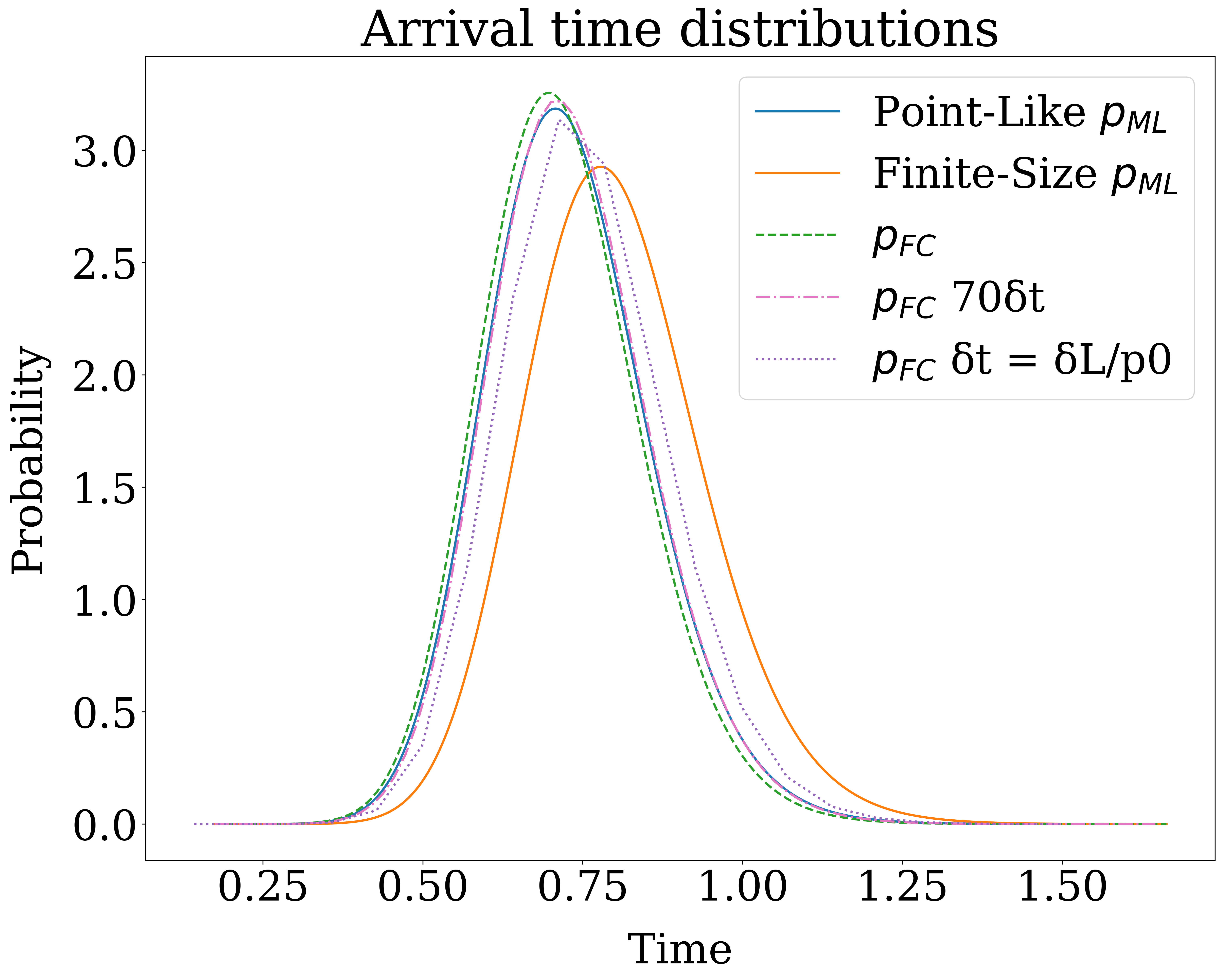}
    \caption{\label{image:MultGaussianWavePacket} Time-of-arrival distributions for the same Gaussian wave packet as in Fig.~\ref{image:GaussianWavePacket}, with $p_{FC}$ evaluated for different time resolutions: $\delta t = \{t_0, 70t_0, \delta L/p_0\}$, as labeled. Increasing $\delta t$ broadens the distribution and shifts it toward later times, reflecting reduced temporal resolution. Even when $\delta t$ is chosen to match the time required for the particle to traverse the finite-size detector ($\delta t = m\delta  L/p_0$ with $m=1$), $p_{FC}$ remains narrower and shifted toward earlier times than the finite-size $p_{ML}$.} 
    \captionsetup{justification=justified}
\end{figure}

From an experimental perspective, it is also relevant to assess the impact of the detector's temporal resolution. As shown in Fig.~\ref{image:MultGaussianWavePacket}, increasing $\delta t$ (i.e., reducing the time resolution) leads to a broadening of the distribution and a shift toward later times. Physically, this reflects the reduced ability of the detector to resolve the arrival time precisely, allowing the particle to propagate further into the detection region before being recorded. \par 

A particularly relevant case is when $\delta t$ is chosen to match the time required for the particle to traverse the detector region, namely $\delta t=\delta L/v_0$, where $v_0=p_0/m$ is the initial average velocity, $\delta L$ is the detector width, and we set $m=1$. In this regime, the coarse temporal resolution implies that the distribution is sampled at relatively few time points, resulting in a less smooth profile with more pronounced transitions. \par  

Even in this regime, $p_{FC}$ remains shifted toward earlier times and exhibits a higher peak amplitude than the finite-size $p_{ML}$. This demonstrates that the conditioning effect persists across different temporal resolutions. \par

Overall, the first-click distributions remain qualitatively distinct from those of the memoryless framework. This confirms that conditioning on the absence of prior clicks plays a significant role in shaping the TOA distribution. In the following section, we investigate whether this behaviour persists for more complex quantum states. \par \medskip

\subsection{Overlapping packets}

In this section, we analyze the TOA distributions for a superposition of two Gaussian wave packets, initially localized at positions $x_0$ and $x_1$, with momenta $p_0$ and $p_1$, respectively. We consider the case $p_1>p_0$, and adjust $x_1$ such that both wave packets arrive at the detector's left boundary with the same average TOA. This corresponds to the faster but initially more distant packet overtaking the slower one at $x=0$, and is ensured by the condition $x_1/p_1=x_0/p_0$. \par

The initial state is taken to be an equally weighted superposition of two Gaussian wave packets  $( |\psi_0 (t)\rangle+|\psi_1 (t)\rangle)/\sqrt{2} $ where, as before, each component $|\psi_k(t)\rangle$ evolves according to Eq.~(\ref{Eq:WaveEvolve}) or Eq.~(\ref{Eq:EvolveU}) depending on the detection employed. \par 

As in the previous section, $p_{ML}$ is computed for both a point-like detector located at the left boundary of $D$ and a finite-size detector covering the entire region. \par

\begin{figure}[t!] 
    \includegraphics[width=1\linewidth]{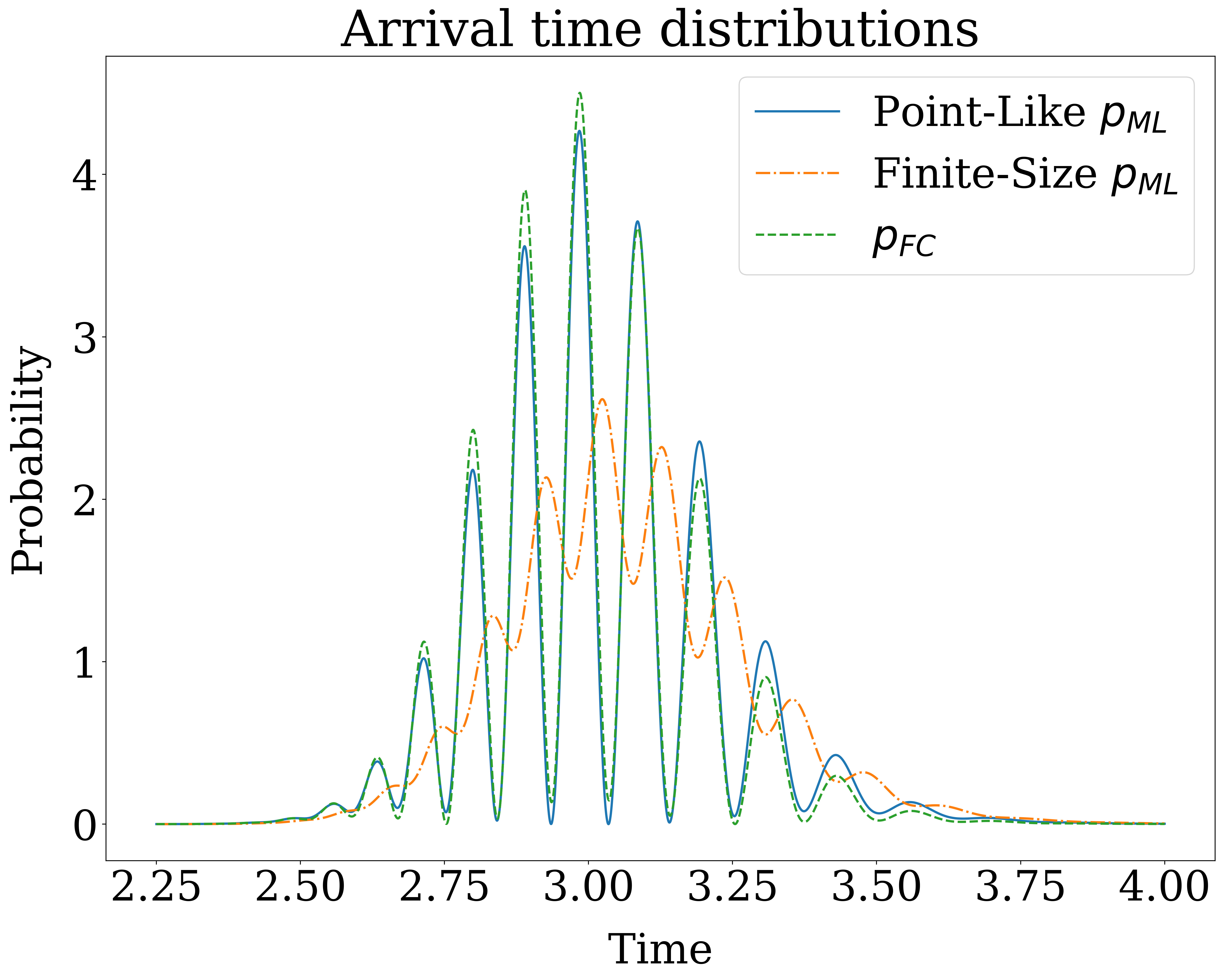}
    \caption{Time of arrival distributions for a superposition of $n=2$ right-moving Gaussian wave packets satisfying the overtaking condition. The parameters of each packet are $x_0 = -30 l_0$, $p_0 =10 \hbar /l_0$ and $p_1 =15 \hbar /l_0$ and $\sigma_0=\sigma_1=1 l_0$, which give $x_1=-45 l_0$, so that both packets arrive simultaneously at the left boundary of the detector, which extends between $a=0$ and $b=l_0$. The distributions display the interference fringe structure characteristic of the two-packet superposition. Compared with the point-like $p_{ML}$, $p_{FC}$ enhances the early peaks and suppresses the later ones, while the finite-size $p_{ML}$ partially smooths the fringes.} 
    \label{image:TwoGaussianpackets}
\end{figure}

The resulting distributions are shown in Fig.~\ref{image:TwoGaussianpackets}. The point-like $p_{ML}$ and $p_{FC}$ both reproduce the interference fringe structure expected from a two-packet superposition. By contrast, the finite-size $p_{ML}$ exhibits a partial smoothing of these fringes due to the extended spatial range of the detector. \par

Compared with the single-packet case, the effect of conditioning is now modulated by interference. Presumably, during destructive interference, when the probability density at the detector is nearly zero, the conditioning barely affects the state because very little amplitude is available to be removed. By contrast, during constructive interference, a significant fraction of the wavefunction is removed at each detection attempt due to the conditioning. \par

Beyond this modulation, the overall effect of conditioning remains consistent with the single-packet scenario: probability is redistributed toward earlier times. Early peaks are enhanced and later peaks are suppressed relative to the point-like $p_{ML}$. This effect is most pronounced near the interference maxima, where the overlap between the two packets produces the largest probability density in the detector region. \par

These results show that the conditioning mechanism is robust in the presence of quantum interference. While the interference pattern shapes the detailed structure of the distribution, it does not alter the qualitative redistribution of probability toward earlier arrival times.

\section{Discussion \label{sec:level4}}

In this work, we introduced a memory mechanism within the Page and Wootters formalism that records the particle’s first detection across successive time intervals defined by the detector's time resolution. This extension yields a well-defined first-click distribution that explicitly accounts for the conditioning on the absence of prior detection.

We analyzed this distribution numerically for two scenarios: a single Gaussian wave packet and a superposition of two overlapping Gaussian wave packets, comparing the results with those of the memoryless framework. In both cases, conditioning on no prior detection progressively removes probability amplitude from the detection region at each measurement attempt. This leads to a redistribution of probability weight toward earlier arrival times, producing distributions that are narrower and exhibit higher peak amplitudes than their memoryless counterparts. This difference cannot be absorbed into a simple normalization, but reflects a genuine modification of the underlying statistics. 

We also showed that the detector's time resolution plays a crucial role in shaping the distribution. Coarser resolutions lead to broader distributions that are shifted toward later times, reflecting the reduced temporal sensitivity and the particle's ability to propagate further into the detector before it clicks. In the two-packet case, the conditioning effect persists in the presence of quantum interference: the redistribution of probability remains, but is modulated by regions of constructive and destructive interference. Importantly, this qualitative difference persists across all scenarios considered and remains robust under changes in the detector’s time resolution.

As expected, a detector which probes the system at successive times necessarily induces a back-action through each null measurement. The absence of a detection event at a given time is itself an interaction with the system and results in a non-trivial update of the system's state. The memoryless framework neglects this effect entirely. Our results demonstrate that this omission is not merely a technical simplification, but leads to qualitatively different predictions for the TOA distributions. In this sense, our results show that detector back-action is not optional, but an essential ingredient for a consistent description of time-of-arrival measurements.

Several directions remain open for future work. A particularly relevant extension is the investigation of the continuous-measurement limit ($\delta t \rightarrow 0$) and its connection with the quantum Zeno effect. In the continuous-measurement limit, projective detection attempts are expected to suppress the arrival probability entirely via the quantum Zeno effect. Understanding how this limit is approached, and whether it can be regularized by replacing projective measurements with weaker detection models, is a natural next step \cite{Barchielli1982,Barchielli1983,PhysRevD.39.2943}.
It would also be interesting to explore the extension to entangled particle pairs and to compare with recent approaches to arrival-time statistics in classical–quantum hybrids ~\cite{kazemi2025arrivaltimeclassical}.

\begin{acknowledgments}
L.C. acknowledges funding from the Horizon Europe project FoQaCiA, GA no.101070558. S.R. acknowledges support from the PRIN MUR Project 2022RATBS4. L.M. acknowledges support from the National Research Centre for HPC, Big Data and Quantum Computing, PNRR MUR Project CN0000013-ICSC. 
\end{acknowledgments}



\bibliography{bib}

@article{MacconeQTM,
  title = {Quantum Measurements of Time},
  author = {Maccone, Lorenzo and Sacha, Krzysztof},
  journal = {Phys. Rev. Lett.},
  volume = {124},
  issue = {11},
  pages = {110402},
  numpages = {6},
  year = {2020},
  month = {Mar},
  publisher = {American Physical Society},
  doi = {10.1103/PhysRevLett.124.110402},
  url = {https://link.aps.org/doi/10.1103/PhysRevLett.124.110402}
}

@article{MacconeQT,
  title = {Quantum time},
  author = {Giovannetti, Vittorio and Lloyd, Seth and Maccone, Lorenzo},
  journal = {Phys. Rev. D},
  volume = {92},
  issue = {4},
  pages = {045033},
  numpages = {9},
  year = {2015},
  month = {Aug},
  publisher = {American Physical Society},
  doi = {10.1103/PhysRevD.92.045033},
  url = {https://link.aps.org/doi/10.1103/PhysRevD.92.045033}
}

@article{Stueckelberg,
  title = {E. C. G. Stueckelberg: A forerunner of modern physics},
  author = {F. Cianfrani, O. M. Lecian},
  journal = {Il Nuovo Cimento B},
  numpages = {10},
  year = {2007},
  month = {2},
  publisher = {Società Italiana di Fisica},
  doi = {http://dx.doi.org/10.1393/ncb/i2007-10359-9}
}

@Inbook{ZEH2009,
author="Zeh, H. Dieter",
editor="Greenberger, Daniel
and Hentschel, Klaus
and Weinert, Friedel",
title="Time in Quantum Theory",
bookTitle="Compendium of Quantum Physics",
year="2009",
publisher="Springer Berlin Heidelberg",
address="Berlin, Heidelberg",
pages="786--792",
isbn="978-3-540-70626-7",
doi="10.1007/978-3-540-70626-7_221",
url="https://doi.org/10.1007/978-3-540-70626-7_221"
}

@article{Rovelli1996,
  title = {Relational quantum mechanics},
  author = {Rovelli, Carlo},
  journal = {International Journal of Theoretical Physics},
  volume = {35},
  issue = {8},
  pages = {1637–1678},
  numpages = {9},
  year = {1996},
  month = {8},
  publisher = {Plenum Pblishing Corporation},
  doi = {10.1007/BF02302261},
  url = {https://doi.org/10.1007/BF02302261}
}

@article{PhysRevD.43.442,
  title = {Time in quantum gravity: An hypothesis},
  author = {Rovelli, Carlo},
  journal = {Phys. Rev. D},
  volume = {43},
  issue = {2},
  pages = {442--456},
  numpages = {0},
  year = {1991},
  month = {Jan},
  publisher = {American Physical Society},
  doi = {10.1103/PhysRevD.43.442},
  url = {https://link.aps.org/doi/10.1103/PhysRevD.43.442}
}

@article{PageandWootters,
  title = {Evolution without evolution: Dynamics described by stationary observables},
  author = {Page, Don N. and Wootters, William K.},
  journal = {Phys. Rev. D},
  volume = {27},
  issue = {12},
  pages = {2885--2892},
  numpages = {0},
  year = {1983},
  month = {Jun},
  publisher = {American Physical Society},
  doi = {10.1103/PhysRevD.27.2885},
  url = {https://link.aps.org/doi/10.1103/PhysRevD.27.2885}
}

@book{4Canadian,
  title={Proceedings of the $4^{th}$ Canadian Conference on General Relativity and Relativistic Astrophysics : University of Winnipeg 16-18 May, 1991},
  author={Relativistic Astrophysics and Gabor Kunstatter and Dwight E. Vincent and Jeff G. Williams},
  year={1992},
  url={https://api.semanticscholar.org/CorpusID:221999738}
}

@article{Cotler_2016,
doi = {10.1088/0031-8949/2016/T168/014004},
url = {https://dx.doi.org/10.1088/0031-8949/2016/T168/014004},
year = {2016},
month = {may},
publisher = {IOP Publishing},
volume = {2016},
number = {T168},
pages = {014004},
author = {Cotler, Jordan and Wilczek, Frank},
title = {Entangled histories},
journal = {Physica Scripta}
}

@book{Yakir
  , author    = "Yakir Aharonov Festschrift"
  , TITLE = "Quantum Theory: A Two-Time Success Story"
  , edition   =	{Struppa, Daniele C. and Tollaksen, Jeffrey M.}
  , publisher = "Springer Milano"
  , YEAR = "2014"
  , doi = {https://doi.org/10.1007/978-88-470-5217-8}
}

@misc{sels2015thermodynamicstime,
      title={The thermodynamics of time}, 
      author={Dries Sels and Michiel Wouters},
      year={2015},
      eprint={1501.05567},
      archivePrefix={arXiv},
      primaryClass={quant-ph},
      url={https://arxiv.org/abs/1501.05567}, 
}

@article{PhysRev.109.571,
  title = {Quantum Limitations of the Measurement of Space-Time Distances},
  author = {Salecker, H. and Wigner, E. P.},
  journal = {Phys. Rev.},
  volume = {109},
  issue = {2},
  pages = {571--577},
  numpages = {10},
  year = {1958},
  month = {Jan},
  publisher = {American Physical Society},
  doi = {10.1103/PhysRev.109.571},
  url = {https://link.aps.org/doi/10.1103/PhysRev.109.571}
}

@book{Feyman
  , author    = "Feynman, Richard P."
  , TITLE = "Simulating physics with computers"
  , journal = {International Journal of Theoretical Physics}
  , publisher = {"Springer Nature SharedIt"}
  , YEAR = "1982"
  , doi = {10.1007/BF02650179}
  , url = {https://doi.org/10.1007/BF02650179}
}

@article{Wootters,
  title = {“Time” replaced by quantum correlations},
  author = {Wootters, William K.},
  journal = {International Journal of Theoretical Physics},
  volume = {23},
  issue = {8},
  pages = {701–711},
  numpages = {11},
  year = {1984},
  month = {Aug},
  publisher = {Springer Nature SharedIt},
  doi = {10.1007/BF02214098},
  url = {https://doi.org/10.1007/BF02214098}
}

@book{vonNeumann+2018,
url = {https://doi.org/10.1515/9781400889921},
title = {Mathematical Foundations of Quantum Mechanics},
author = {John von Neumann},
editor = {Nicholas A. Wheeler},
publisher = {Princeton University Press},
address = {Princeton},
doi = {doi:10.1515/9781400889921},
isbn = {9781400889921},
year = {2018},
lastchecked = {2025-03-19}
}

@article{Jurman_2021,
   title={The time distribution of quantum events},
   volume={396},
   ISSN={0375-9601},
   url={http://dx.doi.org/10.1016/j.physleta.2021.127247},
   DOI={10.1016/j.physleta.2021.127247},
   journal={Physics Letters A},
   publisher={Elsevier BV},
   author={Jurman, Danijel and Nikolić, Hrvoje},
   year={2021},
   month=apr, pages={127247} }

@misc{jurić2022arrivaltimegeneraltheory, 
      title={Arrival time from the general theory of quantum time distributions}, 
      author={Tajron Jurić and Hrvoje Nikolić},
      year={2022},
      eprint={2107.08777},
      archivePrefix={arXiv},
      primaryClass={quant-ph},
      url={https://arxiv.org/abs/2107.08777}, 
}

@article{Page,
  title = {Origin of time asymmetry},
  author = {Hawking, S. W. and Laflamme, R. and Lyons, G. W.},
  journal = {Phys. Rev. D},
  volume = {47},
  issue = {12},
  pages = {5342--5356},
  numpages = {0},
  year = {1993},
  month = {Jun},
  publisher = {American Physical Society},
  doi = {10.1103/PhysRevD.47.5342},
  url = {https://link.aps.org/doi/10.1103/PhysRevD.47.5342}
}

@book{Shankar,
 author = {Shankar, R.},
 doi = {10.1007/978-1-4757-0576-8},
 editor = {},
 pages = {},
 title = {Principles of Quantum Mechanics},
 url = {https://app.dimensions.ai/details/publication/pub.1042261003},
 year = {1994}
}

@misc{simoneroncallo,
      title={Quantum stroboscopy for time measurements}, 
      author={Seth Lloyd and Lorenzo Maccone and Lionel Martellini and Simone Roncallo},
      year={2025},
      eprint={2507.17740},
      archivePrefix={arXiv},
      primaryClass={quant-ph},
      url={https://arxiv.org/abs/2507.17740}, 
}

@article{Roncallo_2023,
   title={When does a particle arrive?},
   volume={7},
   ISSN={2521-327X},
   url={http://dx.doi.org/10.22331/q-2023-03-30-968},
   DOI={10.22331/q-2023-03-30-968},
   journal={Quantum},
   publisher={Verein zur Forderung des Open Access Publizierens in den Quantenwissenschaften},
   author={Roncallo, Simone and Sacha, Krzysztof and Maccone, Lorenzo},
   year={2023},
   month=mar, pages={968} }

@article{Echanobe_2008,
   title={Disclosing hidden information in the quantum Zeno effect: Pulsed measurement of the quantum time of arrival},
   volume={77},
   ISSN={1094-1622},
   url={http://dx.doi.org/10.1103/PhysRevA.77.032112},
   DOI={10.1103/physreva.77.032112},
   number={3},
   journal={Physical Review A},
   publisher={American Physical Society (APS)},
   author={Echanobe, J. and del Campo, A. and Muga, J. G.},
   year={2008},
   month=mar }

@misc{kazemi2025arrivaltimeclassical,
      title={Arrival Time -- Classical Parameter or Quantum Operator?}, 
      author={MohammadJavad Kazemi and MohammadHossein Barati and Ghadir Jafari and S. Shajidul Haque and Saurya Das},
      year={2025},
      eprint={2512.13502},
      archivePrefix={arXiv},
      primaryClass={quant-ph},
      url={https://arxiv.org/abs/2512.13502}, 
}

@article{PhysRevA.64.012501,
  title = {Time-of-arrival distributions for interaction potentials},
  author = {Baute, A. D. and Egusquiza, I. L. and Muga, J. G.},
  journal = {Phys. Rev. A},
  volume = {64},
  issue = {1},
  pages = {012501},
  numpages = {6},
  year = {2001},
  month = {Jun},
  publisher = {American Physical Society},
  doi = {10.1103/PhysRevA.64.012501},
  url = {https://link.aps.org/doi/10.1103/PhysRevA.64.012501}
}

@article{Delgado_1997,
   title={Arrival time in quantum mechanics},
   volume={56},
   ISSN={1094-1622},
   url={http://dx.doi.org/10.1103/PhysRevA.56.3425},
   DOI={10.1103/physreva.56.3425},
   number={5},
   journal={Physical Review A},
   publisher={American Physical Society (APS)},
   author={Delgado, V. and Muga, J. G.},
   year={1997},
   month=nov, pages={3425–3435} }

@article{Grot_1996,
   title={Time of arrival in quantum mechanics},
   volume={54},
   ISSN={1094-1622},
   url={http://dx.doi.org/10.1103/PhysRevA.54.4676},
   DOI={10.1103/physreva.54.4676},
   number={6},
   journal={Physical Review A},
   publisher={American Physical Society (APS)},
   author={Grot, Norbert and Rovelli, Carlo and Tate, Ranjeet S.},
   year={1996},
   month=dec, pages={4676–4690} }

@article{Galapon_2004,
   title={Confined Quantum Time of Arrivals},
   volume={93},
   ISSN={1079-7114},
   url={http://dx.doi.org/10.1103/PhysRevLett.93.180406},
   DOI={10.1103/physrevlett.93.180406},
   number={18},
   journal={Physical Review Letters},
   publisher={American Physical Society (APS)},
   author={Galapon, Eric A. and Caballar, Roland F. and Jr, Ricardo T. Bahague},
   year={2004},
   month=oct }

@article{Anastopoulos_2006, 
   title={Time-of-arrival probabilities and quantum measurements},
   volume={47},
   ISSN={1089-7658},
   url={http://dx.doi.org/10.1063/1.2399085},
   DOI={10.1063/1.2399085},
   number={12},
   journal={Journal of Mathematical Physics},
   publisher={AIP Publishing},
   author={Anastopoulos, Charis and Savvidou, Ntina},
   year={2006},
   month=dec }

@article{Halliwell_2009, 
   title={Quantum arrival time formula from decoherent histories},
   volume={374},
   ISSN={0375-9601},
   url={http://dx.doi.org/10.1016/j.physleta.2009.10.077},
   DOI={10.1016/j.physleta.2009.10.077},
   number={2},
   journal={Physics Letters A},
   publisher={Elsevier BV},
   author={Halliwell, J.J. and Yearsley, J.M.},
   year={2009},
   month=dec, pages={154–157} }

@article{Anastopoulos_2012, 
   title={Time-of-arrival probabilities for general particle detectors},
   volume={86},
   ISSN={1094-1622},
   url={http://dx.doi.org/10.1103/PhysRevA.86.012111},
   DOI={10.1103/physreva.86.012111},
   number={1},
   journal={Physical Review A},
   publisher={American Physical Society (APS)},
   author={Anastopoulos, Charis and Savvidou, Ntina},
   year={2012},
   month=jul }

@article{Vona_2013, 
   title={What Does One Measure When One Measures the Arrival Time of a Quantum Particle?},
   volume={111},
   ISSN={1079-7114},
   url={http://dx.doi.org/10.1103/PhysRevLett.111.220404},
   DOI={10.1103/physrevlett.111.220404},
   number={22},
   journal={Physical Review Letters},
   publisher={American Physical Society (APS)},
   author={Vona, Nicola and Hinrichs, Günter and Dürr, Detlef},
   year={2013},
   month=nov }

@misc{dhar2015quantumtimearrivaldistribution,
      title={Quantum time of arrival distribution in a simple lattice model}, 
      author={Shrabanti Dhar and Subinay Dasgupta and Abhishek Dhar},
      year={2015},
      eprint={1312.5923},
      archivePrefix={arXiv},
      primaryClass={quant-ph},
      url={https://arxiv.org/abs/1312.5923}, 
}

@article{Halliwell_2015,
   title={A self-adjoint arrival time operator inspired by measurement models},
   volume={379},
   ISSN={0375-9601},
   url={http://dx.doi.org/10.1016/j.physleta.2015.07.040},
   DOI={10.1016/j.physleta.2015.07.040},
   number={39},
   journal={Physics Letters A},
   publisher={Elsevier BV},
   author={Halliwell, J.J. and Evaeus, J. and London, J. and Malik, Y.},
   year={2015},
   month=oct, pages={2445–2451} }

@article{Galapon_2018,
   title={Quantizations of the classical time of arrival and their dynamics},
   volume={397},
   ISSN={0003-4916},
   url={http://dx.doi.org/10.1016/j.aop.2018.08.005},
   DOI={10.1016/j.aop.2018.08.005},
   journal={Annals of Physics},
   publisher={Elsevier BV},
   author={Galapon, Eric A. and P. Magadan, John Jaykel},
   year={2018},
   month=oct, pages={278–302} }

@article{PhysRevA.102.053705,
  title = {Quantum-optical implementation of non-Hermitian potentials for asymmetric scattering},
  author = {Ruschhaupt, A. and Kiely, A. and Sim\'on, M. A. and Muga, J. G.},
  journal = {Phys. Rev. A},
  volume = {102},
  issue = {5},
  pages = {053705},
  numpages = {10},
  year = {2020},
  month = {Nov},
  publisher = {American Physical Society},
  doi = {10.1103/PhysRevA.102.053705},
  url = {https://link.aps.org/doi/10.1103/PhysRevA.102.053705}
}

@article{Brunetti2010,
    title = {Time in Quantum Physics: From an External Parameter to an Intrinsic Observable},
    author  = {Brunetti, Romeo and Fredenhagen, Klaus and Hoge, Marc},
    year  = {2010},
    month = oct,
    journal = {Foundations of Physics},
    volume = {40},
    issue = {9},
    ISSN = {1572-9516},
    url = {https://doi.org/10.1007/s10701-009-9400-z},
    doi = {10.1007/s10701-009-9400-z}
}

@article{Aharonov21,
  title = {Measurement of time of arrival in quantum mechanics},
  author = {Aharonov, Y. and Oppenheim, J. and Popescu, S. and Reznik, B. and Unruh, W. G.},
  journal = {Phys. Rev. A},
  volume = {57},
  issue = {6},
  pages = {4130--4139},
  numpages = {0},
  year = {1998},
  month = {Jun},
  publisher = {American Physical Society},
  doi = {10.1103/PhysRevA.57.4130},
  url = {https://link.aps.org/doi/10.1103/PhysRevA.57.4130}
}

@article{Das2019,
    author = {Das, Siddhant and Dürr, Detlef},
    title = {Arrival Time Distributions of Spin-1/2 Particles} ,
    journal = {Scientific Reports} ,
    year = {2019},
    month = feb,
    volume = {9},
    issue = {1},
    ISSN = {2045-2322},
    url = {https://doi.org/10.1038/s41598-018-38261-4},
    doi = {10.1038/s41598-018-38261-4}
}

@article{PhysRevA.58.840, 
  title = {Time of arrival in quantum and Bohmian mechanics},
  author = {Leavens, C. R.},
  journal = {Phys. Rev. A},
  volume = {58},
  issue = {2},
  pages = {840--847},
  numpages = {0},
  year = {1998},
  month = {Aug},
  publisher = {American Physical Society},
  doi = {10.1103/PhysRevA.58.840},
  url = {https://link.aps.org/doi/10.1103/PhysRevA.58.840}
}

@article{GaugeQuantumTime,
    title = {Can We Gauge Quantum Time of Flight?},
    author = {Anil Ananthaswamy},
    journal = {Scientific American},
    year = {2021},
    url ={https://www.scientificamerican.com/article/this-simple-experiment-could-challenge-standard-quantum-theory/}
}

@article{Barchielli1982,
author = {Barchielli, Alberto and Lanz, L. and Prosperi, G.},
year = {1982},
month = {11},
pages = {79-121},
title = {A model for the macroscopic description and continual observations in quantum mechanics},
volume = {72},
journal = {Il Nuovo Cimento B},
doi = {10.1007/BF02894935}
}

@article{Barchielli1983,
    author = {Barchielli, A. and Lanz, L. and Prosperi, G. M.} ,
    title = {Statistics of continuous trajectories in quantum mechanics: Operation-valued stochastic processes} ,
    journal = {Foundations of Physics},
    year = {1983},
    month = aug,
    volume = {13},
    issue = {8},
    ISSN = {1572-9516},
    url = {https://doi.org/10.1007/BF01906270},
    doi = {10.1007/BF01906270}
}

@article{PhysRevD.39.2943,
  title = {Quantum limited detectors for weak classical signals},
  author = {Peres, Asher},
  journal = {Phys. Rev. D},
  volume = {39},
  issue = {10},
  pages = {2943--2950},
  numpages = {0},
  year = {1989},
  month = {May},
  publisher = {American Physical Society},
  doi = {10.1103/PhysRevD.39.2943},
  url = {https://link.aps.org/doi/10.1103/PhysRevD.39.2943}
}

\end{document}